\documentstyle[aps,prl]{revtex}
\input epsf.sty
\begin{document}

\twocolumn[\hsize\textwidth\columnwidth\hsize\csname@twocolumnfalse\endcsname

\draft

\title{Spin-gap effect on resistivity in the t-J model}
\author{Masaru Onoda\cite{onoda}}
\address{Department of Physics, University of Tokyo,
Hongo, Tokyo, 113-0033 Japan
}
\author{Ikuo Ichinose\cite{ichinose}}
\address{Institute of Physics, University of Tokyo,
Komaba, Tokyo, 153-8902 Japan}
\author{Tetsuo Matsui\cite{matsui}}
\address{Department of Physics, Kinki University,
Higashi-Osaka, 577-8502 Japan
}

\date{\today}

\maketitle

\begin{abstract}
We calculate the spin-gap effect on dc resistivity in the t-J model of
high-$T_{\rm c}$ cuprates
by using the Ginzburg-Landau theory coupled with a gauge field
as its effective field
theory to get  $\rho(T) \propto T \{ 1-c\:(T^* -T)^d \}$, where $T^*$
is the spin-gap onset
temperature. By taking the compactness of massive gauge field into account,
the exponent $d$ deviates from its mean-field value $1/2$ and
becomes a nonuniversal
$T$-dependent quantity, which improves the correspondence with the experiments.
\end{abstract}


]

The idea of charge-spin separation (CSS) by Anderson\cite{Anderson} which
accounts for the anomalous behavior of various normal-state properties of 
high-$T_{\rm c}$ cuprates\cite{etc} allows us to treat holons and spinons 
introduced in the slave-boson (SB) mean-field theory (MFT) of the t-J model 
as quasi-free particles. Fluctuations around MFT are described by gauge 
fields coupled to holons and spinons, the effects of which may be calculated 
in perturbation theory. Actually, by using a massless 
gauge field, Nagaosa and Lee \cite{Nagaosa&Lee} obtained 
the dc resistivity $\rho(T) \propto T$,
which agrees with the experiment for $T > T^*$ \cite{Gurvitch&Fiory} where
$T^*$ is the onset temperature of spin gap.

For $T < T^*$, the experimentally observed $\rho(T)$ reduces from this
$T$-linear behavior \cite{Ito&Takenaka&Uchida,Yasuoka}. Because the gauge 
field is expected to acquire a mass $m_A$ in the spin-gap state, this reduction 
could be understood as a mass effect; the fluctuations of the gauge field become 
weaker and the scatterings between holons (the carriers of charge)
and gauge bosons are reduced.
Actually, in the previous paper \cite{Onoda&Ichinose&Matsui},
we obtained the following result \cite{X};
\begin{eqnarray}
\rho(T)
&\simeq& \frac{3 \pi \bar{m}}{2 e^2 n_B} k_{\rm B} T
 \Bigl[ 1+X(T) - \sqrt{\{1+X(T)\}^2 -1} \Bigr],\nonumber\\
X(T) &=& \frac{m_A^2(T)}{8\pi \tilde{n}_B(T)},
\quad \frac1{\bar{m}} = \frac1{m_F} + \frac{f_B(-\mu_B)}{2m_B},
\label{rho}
\end{eqnarray}
where $m_{F(B)}$ is the spinon (holon) mass, 
$f_B(-\mu_B) = [\exp(-\beta\mu_B)-1]^{-1}$,
with the holon chemical potential
$\mu_B$, $n_B$ is the holon density, and $\tilde{n}_B = n_B /f_B(-\mu_B)$.
The factor in the square brackets represents the reduction rate
due to $m_A^2(T)$. By assuming the behavior
$m_A(T) \propto (T^* - T)^{d}$$ (d > 0)$
near $T^*$, and ignoring the weak $T$-dependence in $\bar{m}(T)$ and
$\tilde{n}_B(T)$, we get
\begin{eqnarray}
\rho(T) \propto T [ 1-c\:(T^* -T)^d ].
\end{eqnarray}
The MF value $d = 1/2$ is excluded since the experiment shows smooth 
deviations from the $T$-linear form, which requires $ d > 1$.
In this letter, we calculate $m_A$ by setting up the effective field theory
and taking the compactness of gauge field into account, finding  that $d$ is 
{\it not} a universal constant but  has the $T$-dependence.

The effective theory is the Ginzburg-Landau (GL) theory $L(\lambda, A_i)$ 
coupled with a gauge field $A_i$, where the complex scalar $\lambda$ 
represents the $d$-wave spinon pairing. 
Halperin, Lubensky, and Ma \cite{Halperin&Lubensky&Ma}
considered a similar system in 3D, where
$A_i$ corresponds to the electromagnetic field and $\lambda$ to the
Cooper-pair field. They calculated the effect of $A_i$ on $\lambda$, which  
converted the second-order phase transition to first-order.  
More recently, Ubbens and Lee \cite{Ubbens&Lee} calculated 
the one-loop effect of $A_i$ in the SB MFT of the 2D t-J model, and concluded again 
that the pairing transition at $T^*$  becomes first order.
However, their $T^*$ appears below the superconducting transition temperature 
$T_{\rm c}$, so they concluded that the spin-gap phase 
is completely destroyed by gauge-field fluctuations.
In the present study, we take the compactness of $A_i$ into account, 
which originates from the t-J model defined on the lattice itself 
and gives rise to interactions like $\lambda^2 \cos(2 a A_i)$, 
where $a$ is the lattice constant.
Even in the CSS state, it generates nontrivial vertices 
that are  missing in the usual treatments \cite{Ubbens&Lee} which use 
$\lambda^2  A_i^2$. We find that the periodic interaction
stabilizes the system so as to have a phase transition above $T_{\rm c}$.

For the 2D GL theory coupled with a gauge field,
Nagaosa and Lee \cite{vortex} argued that vortices put 
a phase transition into a crossover.
On the other hand, the 3D system has a genuine phase transition.
This is supported by Monte Carlo simulations 
and other studies \cite{kajantie}. 
In the low-temperature phase (spin-gap phase, Higgs phase),
vortex loops do not proliferate, while in the high-temperature phase,  
they do. The former phase is well described by the usual order 
parameter $\lambda$, while the latter is described by the disorder 
parameter that measures the vortex-loop density.
We assume a small but finite three-dimensionality with anisotropy 
$\alpha > 0$ ($\alpha = 1$ for 3D and $\alpha =0$ for 2D), so there takes 
place a genuine phase transition at $T^*$ in the present model.
 
For a sufficiently small $\alpha$, the 3D critical behavior is observed 
only in the small interval in $T$ near $T^*$ which vanishes 
as $\alpha \rightarrow 0$. Beyond this, 
calculations in the pure 2D system ($\alpha = 0$) 
should give a reliable result according to
the general theory of critical phenomena. 
There is a good example of this; the antiferromagnetic transition of 
cuprates at $T = T_{AF}$ \cite{crossover}, for which the 3D behavior 
takes place for $|T - T_{AF}| \lesssim O(1/|\ln\alpha |^2)$.
Beyond this interval, the 2D results fit the experimental data well.

Let us start with the SB t-J Hamiltonian given by
\begin{eqnarray}
H&=&-t\sum_{x, i,\sigma} \Bigl(b^{\dagger}_{x+ i}f^{\dagger}_{x \sigma}
f_{x+ i\:\sigma}b_x+\mbox{H.c.}\Bigr)\nonumber\\
&&-\frac{J}{2} \sum_{x, i} \Bigl|
f^{\dagger}_{x\uparrow}f^{\dagger}_{x+ i\downarrow}
- f^{\dagger}_{x\downarrow}f^{\dagger}_{x+ i\uparrow}\Bigr|^2 + H_\mu,
\nonumber\\
H_\mu &=& -\sum_{x}\Bigl(\tilde{\mu}_B  b^{\dagger}_{x} b_{x} 
+ \tilde{\mu}_F \sum_{\sigma}
f^{\dagger}_{x\sigma} f_{x\sigma} \Bigr),
\label{Hsb}
\end{eqnarray}
where $f_{x \sigma}$ is the fermionic spinon operator with spin
$\sigma$ ($= \uparrow, \downarrow$) at site $x$ of a 2D lattice, and
$b_x$ is the bosonic holon operator. The direction index $i$ ($= 1, 2$)
is also used as unit vectors. $\tilde{\mu}_{B,F}$ are the chemical potentials
to enforce $\langle b^{\dagger}_{x} b_{x} \rangle = \delta$,
$\sum_{\sigma} \langle   f^{\dagger}_{x\sigma} f_{x\sigma} \rangle = 1-\delta$
where $\delta$ is the doping parameter \cite{A0}.
We introduce the complex auxiliary fields $\chi_{xi}$ and $\lambda_{xi}$
on the link $(x, x+i)$ to decouple both $t$ and $J$ terms as
\begin{eqnarray}
H_{\rm MF}&=&\sum_{x, i}\Bigl[
 \frac{3J}{8}\:|\chi_{xi}|^2+{2\over 3J}\:|\lambda_{xi}|^2 
\nonumber\\
& &-\Bigl\{\chi_{xi}\Bigl(\frac38 J\sum_{\sigma}
f^{\dagger}_{x+i\:\sigma}\:f_{x \sigma}
+t b^{\dagger}_{x+ i}b_x \Bigr)+\mbox{H.c.}\Bigr\} 
\label{Hdec1}\\
& &- \frac12
\Bigl\{ \lambda_{xi}
\Bigl(f^{\dagger}_{x\uparrow}f^{\dagger}_{x+ i\downarrow}
- f^{\dagger}_{x\downarrow}f^{\dagger}_{x+ i\uparrow}\Bigr)+\mbox{H.c.}
\Bigr\}\Bigr] + H_{\mu}.\nonumber
\end{eqnarray}
$\chi_{xi}$ describes hoppings of holons and spinons, while $\lambda_{xi}$
describes the resonating-valence-bond (singlet spin-pair) amplitude.
We shall treat their radial parts as MF's. 

In path integral formalism, the partition function
$Z(\beta) \; [\beta \equiv(k_{\rm B} T)^{-1}]$ is given by
\begin{eqnarray}
Z &=& \int [db][df][d\chi][d\lambda] \: \exp (-S),\nonumber\\
S &=& \int_0^{\beta}d\tau \Bigl[\sum_x \Bigl( \bar{b}_x\dot{b}_x +
\sum_\sigma \bar{f}_{x\sigma}\dot{f}_{x\sigma}\Bigr) + H_{\rm MF}\Bigr],
\end{eqnarray}
where $\tau$ is the imaginary time and
$\dot{b}_x =\partial b_x(\tau)/\partial\tau$,
etc. Let us consider the low-energy effective theory
at temperatures $T \ll T_{\rm CSS}$, where $T_{\rm CSS}$ is the critical
temperature {\it below} which the CSS takes place as 
a deconfinement phenomenon\cite{Ichinose&Matsui}.
It is reasonable to translate the lattice variables to the continuum fields
like $f_{x\sigma} \to a f_\sigma(\mbox{\boldmath $x$})$, etc. using
the lattice constant.
The Hamiltonian of the continuum field theory is given by
\begin{eqnarray}
  H&=&N_{\rm site}\frac34 J\chi^2
  + m_F\:\chi  \int d^2 x\:\Bigl[ |\lambda_{s}(\mbox{\boldmath $x$})|^2
    +|\lambda_{d}(\mbox{\boldmath $x$})|^2 \Bigr]\nonumber\\
  & &+\int d^2 x\: \Bigl[\frac{1}{2m_B}
    \sum_i\Bigl|D_i b(\mbox{\boldmath $x$})\Bigr|^2
     -\mu_B\Bigl|b(\mbox{\boldmath $x$})\Bigr|^2\Bigr]\\
  & &+\int d^2 x\: \Bigl[\frac1{2m_F}\sum_i
    \Bigl|D_i f_{\sigma}(\mbox{\boldmath $x$})\Bigr|^2
     -\mu_F\Bigl|f_{\sigma}(\mbox{\boldmath $x$})\Bigr|^2\Bigr]
\nonumber\\
  & &+\int \frac{ d^2 k\: d^2 q}{(2 \pi)^4} \Bigl[
  \Delta_{\rm SG}(\mbox{\boldmath $k$},\mbox{\boldmath $q$})
  f^{\dagger}_{\uparrow}(\mbox{\boldmath $k$}
  \! +\! \frac{\mbox{\boldmath $q$}}2)
  f^{\dagger}_{\downarrow}(-\mbox{\boldmath $k$}
  \! +\! \frac{\mbox{\boldmath$q$}}2)
  \! +\!\mbox{H.c.} \Bigr], \nonumber
\end{eqnarray}
where $\mu_{B, F}$ are the chemical potentials 
for the continuum theory, 
$f_{\sigma}(\mbox{\boldmath $k$})$ is the Fourier transform of
$f_{\sigma}(\mbox{\boldmath $x$})$, and
\begin{eqnarray}
&&\lambda_{s,d} = \frac12(\lambda_{1}\pm\lambda_{2}),\quad
\frac{1}{2m_B} = t\chi a^2,\nonumber\\
&&\frac{1}{2m_F} = \frac38 J\chi a^2,\quad
k_F^2 = \frac{2\pi}{a^2}(1-\delta),\nonumber\\
&&\Delta_{\rm SG}(\mbox{\boldmath $k$},\mbox{\boldmath $q$})
  = 2(1-\delta)\Bigl(\frac{k^{2}_{1}-k^{2}_{2}}{k^2_F}\Bigr)
  \lambda_{d}(\mbox{\boldmath $q$})
  -2\delta\:\lambda_{s}(\mbox{\boldmath $q$}).
\end{eqnarray}
To obtain $H$, we modified the dispersions of holons and spinons
from cosine form to the quadratic one.
$D_i \equiv \partial_i - i A_i$ is the covariant derivative with
the gauge field $A_i(\mbox{\boldmath $x$})$.
Here we introduced $A_i(\mbox{\boldmath $x$})$ by the correspondence 
$\chi_{xi} \to \chi \exp[i a A_i(\mbox{\boldmath $x$})]$, 
where $\chi$ is the radial part of $\chi_{xi}$.
We ignore the fluctuations of $|\chi_{xi}|$ since $T \ll T_{\chi}$, 
where $T_{\chi}$ is the onset 
temperature of $\chi$. In the SB MFT, $\chi$ is estimated at small 
$\delta$'s as
\begin{eqnarray}
\chi
&\simeq& \Bigl\langle\sum_{\sigma}f^{\dagger}_{x+i\: \sigma}\:f_{x \sigma}
+ \frac{8t}{3J}b^{\dagger}_{x+i}b_x \Bigr\rangle_{\rm MF}\nonumber\\
&\simeq& \frac{4}{\pi^2}\sin^2
\Bigl(\frac{\pi}2\sqrt{1-\delta}\Bigr) 
+ \frac{8t}{3J}\delta,
\end{eqnarray}
if the spinon pairing $\lambda_{xi}$ is neglected.

To obtain the effective action of $\lambda_i$ and $A_i$,
$b$ and $f_{\sigma}$ are integrated by the standard bilinear integrations.
This procedure generates dissipative terms of $\lambda_i$ and $A_i$.
The most singular contributions to $Z$ from the integrations over
$\lambda_i$ and $A_i$ come from their static ($\tau$-independent) modes,
so we keep only the static modes in the effective
Lagrangian density, which is given up to the fourth-order
in fields and derivatives by
\begin{eqnarray}
L_{\rm eff}&=&
a_s |\lambda_{s}|^2 + a_d |\lambda_{d}|^2
+\:4 b\: \delta^4|\lambda_{s}|^4
+\frac{3}{2} b\ (1-\delta)^4|\lambda_{d}|^4
\nonumber\\
&&+\: 2 b\:\delta^2(1-\delta)^2
\Bigl(4|\lambda_{s}|^2
|\lambda_{d}|^2 + \bar{\lambda}_{s}^2
\lambda_{d}^2
+ \bar{\lambda}_d^2
\lambda_{s}^2 \Bigr)\nonumber\\
&&+\:  c\:\sum_i\Bigl(
2\delta^2|{\cal D}_{i}\lambda_{s} |^2
+(1-\delta)^2|{\cal D}_{i}\lambda_{d} |^2
\Bigr)   \nonumber\\
&&+\: c\: \delta(1-\delta) \Bigl(
\overline{{\cal D}_{1}\lambda_{s}}
{\cal D}_{1}\lambda_{d}
-\overline{{\cal D}_{2}\lambda_{s}}
{\cal D}_{2}\lambda_{d} + \mbox{H.c.}
\Bigr)  \nonumber\\
&&+\: \frac1{12\pi\bar{m}}
\sum_{i j}\frac14 F_{ij} F_{ij},
\end{eqnarray}
where $\bar{m}$ is defined in Eq.(\ref{rho}), 
and
\begin{eqnarray}
a_s &=& m_F\chi-\frac{2}{\pi}m_F\delta^2
\ln\Bigl(\frac{2e^{\gamma}}{\pi}\beta\omega_\lambda\Bigr),\nonumber\\
a_d &=&
m_F\chi - \frac{m_F}{\pi}(1- \delta)^2\ln\Bigl(\frac{2e^{\gamma}}{\pi}
\beta\omega_\lambda\Bigr),
\nonumber\\
b &=&  \frac{m_F}{\pi}\frac{7\zeta(3)}{8\pi^{2}}\beta^2,
\quad c = \frac{k_F^2}{4\pi m_F}
\frac{7\zeta(3)}{8\pi^{2}} \beta^2,\nonumber\\
{\cal D}_{i} &=& \partial_{i}-2iA_{i}, \quad
F_{ij} = \partial_{i}A_{j} - \partial_{j}A_{i},
\end{eqnarray}
where $\gamma$ is the Euler number.   
$\omega_\lambda$ is the cutoff of the spinon  energy 
[$\xi \equiv k^2/(2m_F) - \mu_F$] in the one-loop integrals 
representing spinon pairings,
and is estimated as $\omega_\lambda \sim O(\mu_F)$.
From the potential energy of $\lambda_i(\mbox{\boldmath $x$})$,
the system favors the d-wave state
at small $\delta$'s, and the s-wave state at large $\delta$'s.
Let us focus on small $\delta$'s by parameterizing
$\lambda_1(\mbox{\boldmath $x$})=
\lambda \exp[ i\theta(\mbox{\boldmath $x$})], \
\lambda_2(\mbox{\boldmath $x$})=
-\lambda \exp[i\theta(\mbox{\boldmath $x$})]$. 
Here we introduced $\lambda$,
the spin-gap amplitude, for the radial parts of 
$\lambda_{i}(\mbox{\boldmath $x$})$, ignoring their fluctuations.
Then we have the  effective Lagrangian density,
\begin{eqnarray}
L_{\rm eff} & = & L_{\lambda} + L_A, \nonumber\\
L_{\lambda}& =&
a_d  \lambda^2 + \frac{3}{2} b\: (1-\delta)^4\lambda^4,\\
L_A & =&  c\: (1-\delta)^2\lambda^2
\Bigl(\partial_{i}\theta - 2A_{i}\Bigr)^2   + \frac1{12\pi\bar{m}}
\sum_{i j}\frac14 F_{ij} F_{ij}.
\nonumber
\label{Leff}
\end{eqnarray}
From the above $L_{\lambda}$, the MF result is obtained as
\begin{eqnarray}
&&k_{\rm B} T_{\lambda}
=\frac{2e^{\gamma}}{\pi}\: \omega_\lambda
\exp\Bigl[-\frac{\pi\chi}{(1-\delta)^2}\Bigr],
\nonumber\\
&&\frac{7\zeta(3)}{8\pi^{2}}\: (1-\delta)^2
\left[\frac{\lambda(T)}{k_{\rm B}T}\right]^2
\simeq \frac{1}{3}
\Bigl(1-\frac{T}{T_{\lambda}}\Bigr),
\label{MFT}
\end{eqnarray}
where $T_{\lambda}$ is the critical temperature below 
which $\lambda$ develops, 
and should be identified as $T^*$.
The second result is reliable at $T$ near $T_\lambda$.

Let us consider the effect of $L_A$ on $L_{\lambda}$ 
by integrating over $A_i$.
We have treated $A_{i}$ as a noncompact gauge field
although it was originally compact.
This procedure is appropriate for the kinetic term of $A_{i}$,
because we consider the region $T\ll T_{\rm CSS}$.
We will respect the compactness of $A_{i}$ and the angle-nature of 
$\theta$ by considering the following new Lagrangian $L_B$ 
with the periodic mass term;  
\begin{eqnarray}
L_B &=&\frac{1}{12\pi\bar{m}}
\sum_{i j}\frac14 F_{ij} F_{ij}
\nonumber\\
&&+ \: c\: (1-\delta)^2\lambda^2 \cdot
\frac1{a^2}
\Bigl[4-\sum_{i}2\cos\Bigl(2 a B_{i}\Bigr)\Bigr],
\end{eqnarray}
where we introduced 
the Proca (massive vector) field $B_i \equiv A_i - \partial_i\theta/2$. 
($F_{ij}=\partial_i B_j - \partial_j B_i$.) Let us
take the unitary gauge. Then the integrals reduce 
as $[d\theta][dA_i] \equiv [d\theta][dB_i] \rightarrow [dB_i]$.

Let us estimate the gauge-field mass by
the variational method.
We choose the variational Lagrangian $L_B'$ for $L_B$  as
\begin{eqnarray}
L_B' &=&\frac{1}{12\pi\bar{m}} \Bigl(
\sum_{i j}\frac14 F_{ij} F_{ij}
+\sum_i\frac{m^2_{A}}2\: B_i B_i
\Bigr),
\end{eqnarray}
where $m_{A}$ is a variational parameter.
The variational free energy density
$F_B = F_B' +\langle L_B - L_B'
\rangle'$is given by
\begin{eqnarray}
F_B(m_A)
&=& F_B(0) + \frac{k_{\rm B} T}{8\pi}\:m^2_{A}
-\frac{4 c\: (1-\delta)^2\lambda^2}{a^2}
\Bigl(\frac{m^2_A}{q^2_c}\Bigr)^{\frac{T}{\Theta(T)}},
\nonumber\\
k_{\rm B}\Theta(T)&\equiv&
\frac1{3 a^2 \bar{m}}=\chi\Bigl[\frac{J}4+\frac{t}3\:f_B(-\mu_B)\Bigr],
\label{freeenergy}
\end{eqnarray}
where $q_c$, the momentum cutoff of $B_i$, is $O(a^{-1})$.
We have omitted the higher-order terms of $O( m^4_A/q^4_c)$.
Note that we took the definition of the propagator at the same point and
the trace of a functional operator $\hat{O}$ as
\begin{eqnarray}
\langle B_i(\mbox{\boldmath $x$}) B_j(\mbox{\boldmath $x$})\rangle
&\equiv&\lim_{\mbox{\boldmath $y$}\to \mbox{\boldmath $x$}}
\langle B_i(\mbox{\boldmath $x$}) B_j(\mbox{\boldmath $y$})\rangle,
\nonumber\\
\mbox{Tr}\: \hat{O}
&\equiv&\int d^2 x \lim_{\mbox{\boldmath $y$}\to \mbox{\boldmath $x$}}
\langle \mbox{\boldmath $x$}| \hat{O} | \mbox{\boldmath $y$}\rangle,
\end{eqnarray}
By minimizing $F_B(m_A)$, we get
\begin{eqnarray}
m^2_A (T)& = &
q^2_c\Bigl[\frac{96\pi \bar{m}}{q_c^2}
\:c\:(1-\delta)^2\lambda^2\Bigr]^{2 d(T)}, \nonumber\\
d(T) &=&\frac{\Theta(T)}{2[\Theta(T)-T]}.
\end{eqnarray}
We note that the fluctuations of $A_i$ do not affect
the MF result (\ref{MFT}) as long as $d(T) > 1$
since then the order of the corrections becomes higher than $\lambda^4$.
Thus the gauge-field mass $m_A(T)$ starts to develop continuously at
$T_\lambda$ as $m_A(T) \propto (T_{\lambda} - T)^{d(T_{\lambda})}$.
That is, the exponent $d$ is neither $1/2$ nor a constant,
and drastically changes especially when $T_\lambda \sim T_A$, where $T_A$ 
is a root of the equation $T_A=\Theta(T_A)$, at which $d(T)$ diverges.
This is in strong contrast with the noncompact case \cite{Ubbens&Lee}. 
 
If we write $q_c^2 = \epsilon a^{-2}$, we have 
$m_A^2(T) = q_c^2 z\: (1-T/T_{\lambda})^{2d(T)}$ with  
$z = [16\pi(1-\delta)\bar{m}/(\epsilon m_F)]^{2d(T)}$.
A straightforward  estimation,
$\pi q_c^2 = (2\pi/a)^2$, by keeping the area of momentum space, 
gives $\epsilon = 4\pi$. However, this gives rise to a nonrealistic 
curve of $\rho(T)$ that deviates from the $T$-linear behavior
and decreases too rapidly, 
due to the large factor $z \simeq 4^{2d(T)}$.
Thus we regard $q_c$ to be a parameter of the effective theory,
and choose $\epsilon$ so as to obtain a reasonable $\rho(T)$.
For example, we require $z=1$, which implies 
$\epsilon = 16\pi(1-\delta)\bar{m}/m_F$. 

Finally, we need to consider  the renormalization effect of
the hopping parameter $t$.
We assume that the 3D system exhibits  
Bose condensation at the temperature scale of 
$T_{B} \simeq  2\pi n_B/m_B = 4\pi t \chi \delta$, and regard
$T_{B}$ to be the observed  
$T_{\rm c}$ in the lightly-doped region. Since $t \sim 0.3$ eV gives 
rise to $T_B \sim 3000$ K at $\delta \sim 0.15$,
one needs to use an effective $t^* \sim 0.01$ eV  in place of $t$ so 
as to obtain a realistic $T_{\rm c} \sim 100 $K \cite{tstar}.

\begin{figure}
  \begin{picture}(200,120)
    \put(0,0){\epsfxsize 200pt \epsfbox{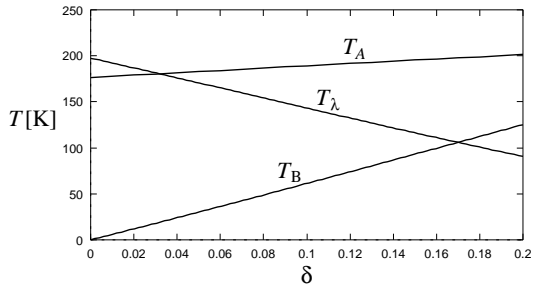}}
  \end{picture}
  \caption{
    Mean-field phase diagram of the t-J model.
    $T_B$ is the Bose condensation temperature.
    $T_\lambda$ is the spin-gap onset temperature. 
    $T_A$ is the root of $T=\Theta(T)$ at which $d(T)$
    diverges. We chose
    $t^*=0.01$ eV, $J = 0.15$ eV, and 
    $\omega_\lambda = \pi J \chi/(2e^{\gamma})$.
    }
\label{phase}
\end{figure}

\begin{figure}
  \begin{picture}(200,200)
    \put(0,0){\epsfxsize 200pt \epsfbox{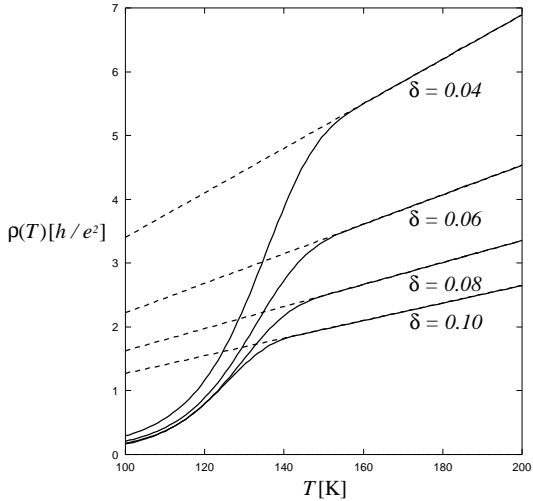}}
  \end{picture}
  \caption{
    Resistivity $\rho(T)$ in $h/e^2$ for several $\delta$'s   
with the parameters chosen in Fig.1. 
The dotted lines represent the case of $X(T)  = 0$ in (1).
The exponent $d(T_{\lambda})$ decreases as 16.4, 4.8,  2.8,  2.0, 
as $\delta$ increases. 
}
\label{resistivity}
\end{figure}

We show in Fig.\ref{phase} the phase diagram with
the spin-gap on-set temperature $T_{\lambda}$,
$T_A$ at which the mass exponent $d$ diverges, and $T_B$.
In Fig.\ref{resistivity}, we plot  $\rho(T)$.
As explained, the curves reproduce the experimental data
much better than those with $d = 1/2$ of the MF result, showing  smooth
departures from the $T$-linear curves, i.e., $d (T_\lambda) > 1$
for the region of interest, $0.05 \lesssim \delta \lesssim 0.15$. 

The present results of $\rho(T)$ support that our treatment of
gauge-field fluctuations by the variational treatment of compactness
is suitable to describe the spin-gap state in the t-J model,
although more investigation is certainly necessary.



\acknowledgments
M.~Onoda is financially supported by Research Fellowships 
of the Japan Society for the Promotion of Science for Young Scientists.



\begin{references}

\bibitem[*]{onoda} Electronic address: onoda@cms.phys.s.u-tokyo.ac.jp
\bibitem[\dagger]{ichinose} Electronic address: ikuo@hep1.c.u-tokyo.ac.jp
\bibitem[\ddagger]{matsui} Electronic address: matsui@phys.kindai.ac.jp

\bibitem{Anderson}
P.~W.~Anderson, Phys. Rev. Lett. {\bf 64}, 1839 (1990).
%
\bibitem{etc}
See, {\it e.g.}, T.~Nishikawa {\it et al.}, J. Phys. Soc. Jpn. {\bf 63},
1441 (1994); J.~Takeda {\it et al.}, Physica C {\bf 231}, 293 (1994);
H.~Y.~Hwang {\it et al.}, Phys. Rev. Lett. {\bf 72}, 2636 (1994).
%
\bibitem{Nagaosa&Lee}
N.~Nagaosa and P.~A.~Lee, Phys. Rev. Lett. {\bf 64}, 2450 (1990);
Phys. Rev. B {\bf 43}, 1233 (1991);
P.~A.~Lee and N.~Nagaosa, Phys. Rev. B {\bf 46}, 5621 (1992).
See also L.~Ioffe and P.~Wiegmann, Phys. Rev. Lett. {\bf 65}, 653 (1990);
I.~Ioffe and G.~Kotliar, Phys. Rev. B {\bf 42} 10348 (1990).
%
\bibitem{Gurvitch&Fiory}
M.~Gurvitch and A.~T.~Fiory, Phys. Rev. Lett. {\bf 59}, 1337 (1987).
%
\bibitem{Ito&Takenaka&Uchida}
T.~Ito, K.~Takenaka, and S.~Uchida, Phys. Rev. Lett. {\bf 70}, 3995 (1993);
B.~Bucher {\it et al.}, Phys. Rev. Lett. {\bf 70}, 2012 (1993).
%
\bibitem{Yasuoka}
H.~Yasuoka {\it et al.},"Strong Correlation and Superconductivity", ed.
H.~Fukuyama {\it et al.} (Springer Series, Berlin, 1989) p.254;
M.~Takigawa {\it et al.}, Phys. Rev. B {\bf 43}, 247 (1991);
J.~Rossat-Mignot {\it et al.}, Physica C {\bf 185-189}, 86 (1991);
J.~M.~Tranquada {\it et al.}, Phys. Rev. B {\bf 46}, 5561 (1992).
%
\bibitem{Onoda&Ichinose&Matsui}
M.~Onoda, I.~Ichinose, and T.~Matsui, J. Phys. Soc. Jpn. {\bf 67}, 2606 (1998).
%
\bibitem{X}
Eq.(18) of \cite{Onoda&Ichinose&Matsui} used the expression 
$X(T) = 3\bar{m}\: n_F^{\rm S}$ $/(2 m_F \tilde{n}_B)$ where
$n_F^{\rm S} \propto n_F \beta^2 (1-\delta)^2 \lambda^2$.
We obtained it by rewriting $X(T)$ of (\ref{rho}) by
assuming the relation $m_A^2  \propto n_F^{\rm S}$ of the MFT.
Because we shall study the case in which $m_A$ and $n_F^{\rm S}$ are
independent, we cite the original expression (\ref{rho}) here.
%
\bibitem{Halperin&Lubensky&Ma}
B.~I.~Halperin, T.~C.~Lubensky, and S.~Ma, Phys. Rev. Lett. {\bf 32}, 292
(1974).
See also S.~Coleman and E.~Weinberg, Phys. Rev. D {\bf 7}, 1888 (1973).
%
\bibitem{Ubbens&Lee}
M.~U.~Ubbens and P.~A.~Lee, Phys. Rev. B {\bf 49}, 6853 (1994).
%
\bibitem{vortex}
N.~Nagaosa and P.~A.~Lee, Phys. Rev. B {\bf 45}, 966 (1992).
%
\bibitem{kajantie}
K.~Kajantie {\it et al.}, Nucl. Phys. B {\bf 520}, 345 (1998);
Phys. Rev. B {\bf 57}, 3011(1998), and references cited therein. 
See also H.~Kleinert, Lett. Nuovo Cimento {\bf 35}, 405 (1982).
%
\bibitem{crossover}
S.~Chakravarty, B.~I.~Halperin and D.~R.~Nelson,
Phys. Rev. Lett. {\bf 60},1057(1988); Phys. Rev. B {\bf 39}, 2344 (1989);
H.~Yamamoto {\it et al.}, Phys. Rev. B {\bf 44}, 7654 (1991).
%
\bibitem{A0} 
The SB local constraint can be incorporated by inserting
the time-component $A_0$ of gauge field.
However, its fluctuation effects are negligible at $T < T_{\rm CSS}$
since they are short-ranged.
%
\bibitem{Ichinose&Matsui}
I.~Ichinose and T.~Matsui,
Nucl. Phys. B {\bf 394}, 281 (1993); Phys. Rev. B {\bf 51}, 11860 (1995).
N.~Nagaosa, Phys. Rev. Lett. {\bf 71}, 4210 (1993), also studied
a confinement-deconfinement transition of certain dissipative gauge theory 
of fermions. However, the deconfinement occurs there {\it above} the critical
temperature.
%
%
\bibitem{tstar}
F.C. Zhang, C. Gross, T.M. Rice, and H. Shiba,
Supercond. Sci. Technol. {\bf 1}, 36 (1988).
\end{references}
\end{document}